\documentclass[sn-nature]{sn-jnl}

\usepackage{graphicx}%
\usepackage{multirow}%
\usepackage{amsmath,amssymb,amsfonts}%
\usepackage{amsthm}%
\usepackage{mathrsfs}%
\usepackage[title]{appendix}%
\usepackage{xcolor}%
\usepackage{textcomp}%
\usepackage{manyfoot}%
\usepackage{booktabs}%
\usepackage{algorithm}%
\usepackage{algorithmicx}%
\usepackage{algpseudocode}%
\usepackage{listings}%
\usepackage{tikz}
\usepackage{cleveref}

\begin{document}

\title[Chapter Title]{Exploring Dense Crowd Dynamics: State of the Art and Emerging Paradigms}


\author[1]{\fnm{Thomas} \sur{Chatagnon}}\email{t.chatagnon@fz-juelich.de}
\equalcont{These authors contributed equally to this work.}
\author[2]{\fnm{Antoine} \sur{Tordeux}}\email{tordeux@uni-wuppertal.de}
\equalcont{These authors contributed equally to this work.}
\author*[1]{\fnm{Mohcine} \sur{Chraibi}}\email{m.chraibi@fz-juelich.de}
\equalcont{These authors contributed equally to this work.}
\affil[1]{\orgdiv{Institute for Advanced Simulation}, \orgname{Forschungszentrum J\"{u}lich}, \orgaddress{\street{Wilhelm-Johnen-Stra\ss{}e}, \city{J\"ulich}, \postcode{52428}, \country{Germany}}}
\affil[2]{\orgdiv{School for Mechanical Engineering and Safety Engineering}, \orgname{University of Wuppertal}, \orgaddress{\street{Gaußstraße 20}, \city{Wuppertal}, \postcode{42119}, \country{Germany}}}


\abstract{Dense pedestrian crowds may pose significant safety risks, yet their underlying dynamics remain insufficiently understood to reliably prevent accidents. 
In these environments, physical interactions and contact forces fundamentally shape the dynamics of the crowd. 
However, accurately describing these interindividual interactions requires specific modeling and analytical approaches.
This chapter reviews paradigms and models used to represent pedestrian dynamics in various contexts, highlighting the transition from classical approaches to models tailored for dense crowd conditions.
We argue that further investigation is needed, featuring new experimental studies and new modeling paradigms, to better capture the complex dynamics that emerge in high-density situations.}

\keywords{Dense Crowd, Pedestrian Model, Physical Interactions, Biomechanics}



\maketitle


\section{Introduction}

In our increasingly urbanized societies, human crowd management faces crucial challenges.
Large-scale events, if not properly controlled, can lead to discomfort, injuries, or even fatalities \cite{feliciani2022introduction,feliciani2023trends}.
While such incidents may arise from a range of factors, this chapter focuses specifically on issues related to crowd motion stemming from inter-individual physical interactions.
We define \emph{dense crowds} here as tightly packed formations of individuals in which the free motion of each individual is limited and likely to result in physical interactions with others. Such a definition still allows for an infinite variety of crowd configurations which all share a common feature; a significant existence of physical interactions between individuals.
In this situation, contact forces can accumulate and people in the crowd can experience unpleasant or even dangerous body compression. This phenomenon can cause anxiety, fainting, and even asphyxia \cite{Sieben_2023, Lee_2005, feliciani2023trends}, either due to pressure on the upper body or poor air quality at the bottom of the crowd \cite{Sieben_2023}.
Physical interactions can also act as external perturbations that throw people off balance. This loss of balance may then lead to additional restrictions on neighbors' bodies, potentially developing into dangerous collective falls, commonly referred to as crowd collapses \cite{Zhou_2017} or wave motions \cite{Sieben_2023}.
Witness statements from the 2010 Love Parade tragedy contained 32 references to falling, as analyzed by Sieben et al. \cite{Sieben_2023}. These accounts frequently mentioned problems with the legs and feet as well.

Fortunately, most dense crowds do not result in accidents. Large-scale events are constantly occurring around the world, with only a small fraction leading to accidents. This raises a fundamental question: How do we identify when the risks to individuals within a dense crowd become critical?
Understanding and controlling crowd dynamics is not straightforward. Pedestrian crowds are many-body systems of self-driven agents with intricate social and physical interactions. 
So far, this human crowd dynamics has been studied using different kinds of modeling approach. Classical pedestrian models are heuristics based on a superposition of pairwise interactions with neighbors. Two-body interaction models can relatively accurately describe most of the collective dynamics of pedestrians for low and intermediate densities in simple scenarios where behaviors result mainly from social proxemics and anticipatory collision avoidance behavior. 
However, these models are limited for modeling dense crowds in confined spaces. 
In fact, the nature of the mechanisms that govern crowd dynamics changes radically in dense situations where pedestrians begin to have direct contact. 
The interactions are dominated by physical forces due to hard body exclusion and the forces developed by pedestrians to remain standing. 
These are no longer social proxemics or collision avoidance techniques, but physical interactions. There can accumulate, leading to extremely high-pressure forces when hundreds of pedestrians are in contact, and potentially to injuries and fatalities. 

Due to their compact nature, dense crowds are also particularly difficult to study. Understanding the position of the body, the different forces at stake, and the psychology of individuals in such contexts remains an open research topic. In addition, the trigger factors for many accidents in dense crowds remain unclear \cite{Bokhary_1993,Lee_2005,Zhen_2008,Balsari_2017}, further blurring the line between life-threatening and regular densely crowded situations. 
The study \cite{Gu2025} observes that in extremely dense crowds, people do not move randomly but rather exhibit collective back-and-forth oscillations. These oscillations arise naturally due to the physical constraints and interactions between individuals.
The authors propose a first-principles-based theory that explains why crowds at extreme densities spontaneously organize into oscillatory motion.
Despite extensive efforts to model pedestrian behavior over the past thirty years, a significant gap remains in our ability to accurately simulate both high-density, physically interacting crowds and low-density, collision avoidance scenarios. Dense crowds introduce complex emergent phenomena, such as force propagation, oscillating motions, and jamming, that arise from detailed physical interactions among individuals. However, many existing models are tailored to capture these emerging effects \cite{ma2013new,bottinelli2016emergent,Gu2025} or to manage anticipatory \cite{Lu2020a,Xu2021,Hu2023}, collision avoidance \cite{Arul2021,Douthwaite2019} behaviors typical of sparser environments.

In this chapter, we begin by reviewing the current state-of-the-art in pedestrian dynamics modeling, focusing on both fundamental body representations and the modeling paradigms at microscopic and macroscopic scales. 
The limitations of these models in capturing the complex physical interactions that emerge in dense crowds, such as wedging, balance recovery, and force propagation, are eventually discussed.
We then examine insights from biomechanics and interdisciplinary experimental studies that have been developed specifically to understand physical interaction and balance recovery of individuals in dense crowds. 
Throughout this chapter, we outline emerging methodologies and highlight key research gaps, offering perspectives on the necessary advances to develop more accurate models to represent dense crowd dynamics.

\section{Overview of Current Pedestrian Dynamics Models}
In this section we discuss spatial representations of pedestrians and review existing models of pedestrian motion. 
Current models are broadly categorized into two types: microscopic models, which describe the trajectories of individual pedestrians, and macroscopic models, which focus on the evolution of aggregated quantities such as crowd density and flow volume. 

\begin{figure}
    \centering
    \includegraphics[width=\linewidth]{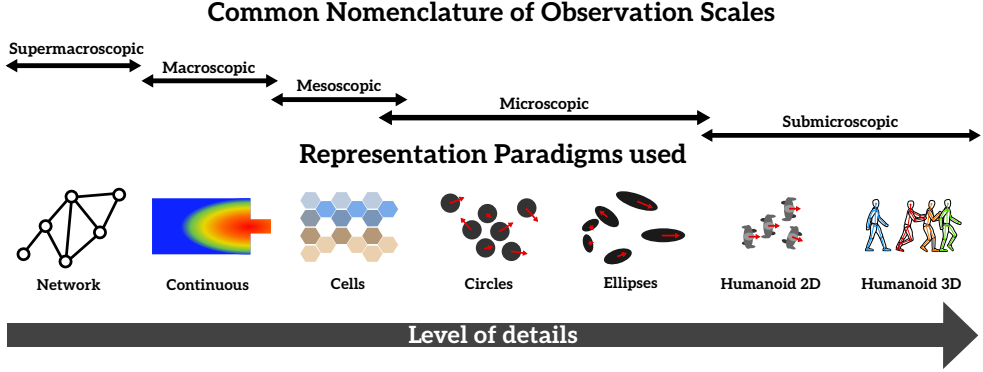}
    \caption{Illustration of representation paradigms for pedestrians in crowd simulation. Different representations are associated with different scales of observation. As more details are needed to capture interactions at lower scales, simulated pedestrians need to get closer to a human body-like representation.}
    \label{fig:representation_paradigms_spectrum}
\end{figure}

We also highlight the limitations of these models, particularly in the context of dense crowds, where physical interactions and complex biomechanical body aspects become more pronounced. 
Finally, we propose viewing spatial representations of pedestrians as a spectrum, with varying levels of detail suited to different applications. 
This conceptual spectrum is illustrated in \Cref{fig:representation_paradigms_spectrum}.

\subsection{Body Shape of Simulated Agents}

A critical factor in enabling a seamless transition between low- and high-density crowd modeling is the accurate representation of the shape of the pedestrian's body. 
In low-density regimes, simplified 2D shapes, such as circles or ellipses, are often sufficient to capture anticipatory behaviors and collision-avoidance strategies.
These approximations offer a computationally efficient means of modeling volume exclusion.
However, such a 2D approximation of the body falls short in high-density scenarios, where the physical interactions between individuals become highly complex.
In these conditions, phenomena such as falling \cite{Sieben_2023}, balance recovery \cite{Jensen_2023,Chatagnon_2024}, and the distribution of contact forces require a three-dimensional representation of the human body.

The 3D modeling of pedestrian shapes unlocks new possibilities by allowing for a more realistic depiction of how forces are transmitted and dissipated across the body.
This level of detail is essential for capturing the emergent dynamics observed in dense crowds, such as oscillatory motion \cite{Gu2025}.

The need for better body representation becomes particularly evident in studies such as those of Kim et al. \cite{Kim2015}, who developed a hybrid approach combining forces based on physics with navigation constraints to simulate crowd behavior in dense scenarios. 
Despite its sophistication, the model's reliance on 2D representations limits its ability to capture the full spectrum of physical interactions typical in dense crowds,  such as pushing with different body parts, dynamic balance recovery, and fall-avoidance strategies.

We argue that these phenomena cannot be accurately reproduced without a detailed three-dimensional representation of the human body.
This paradigm shift, from simplified disk-based approximations to a detailed, multiscale approach that integrates both macroscopic and microscopic interactions, directly addresses the shortcomings of existing models. 
It acknowledges that, while dense crowd conditions introduce unique physical complexities, the behavioral foundations of pedestrian motion remain relevant across all density regimes.
Ensuring the validity of pedestrian models across the full spectrum of densities is not only beneficial, but necessary to create truly predictive and practical tools in pedestrian dynamics research.

One of the first questions that arises when developing microscopic pedestrian models is how to represent the three-dimensional human body within a two-dimensional space without losing critical information.
In industries such as gaming and film, fully three-dimensional body models are essential for producing realistic animations.
However, in most pedestrian dynamics applications, particularly these focused on crowd safety and evacuation, reducing the problem to two dimensions is a common and pratcical choice.
This simplification significantly reduces computational costs and enables a focus on horizontal crowd movements, which are often of primary interest.
Depending on the specific case study (for example, flow through bottleneck or lane formation), this simplified Ansatz often proves sufficient to gain insights into the underlying phenomena.

Despite the advantages of working in two dimensions, researchers have long recognized that overly simplistic body representations can limit the accuracy of pedestrian models.
Early studies often reduced individuals to point-like representation, as in Helbing's work \cite{helbing1995social} or circular disks with fixed radius \cite{Parisi2009} and more recently with variable radii in \cite{Martin2024}. 
These disk-based models offer computational efficiency, particularly in calculating collision avoidance and repulsive forces, since the distance between two pedestrians can be derived directly from their center coordinates. 
However, the rotational symmetry of circles implies that they occupy the same amount of space in all directions, which can be overly restrictive in simulating realistic two-dimensional movement. 
In contrast, pedestrians navigating real environments often turn their bodies sideways to squeeze through narrow gaps or adapt their effectively occupied space depending on their speed and gait, behaviors that  cannot be captured by a uniform circular model.

To address these limitations, researchers introduced ellipses as a more flexible approximation of the human body's 2D projection.
Fruin's early concept of the ``body ellipse'' \cite{Fruin1971}, later expended upon by Templer \cite{Templer1992} demonstrated how ellipses could more accurately reflect the dynamics of human bodies interacting in space.
Building on these ideas, subsequent models presented in \cite{chraibi2011force,Xu2019}, enabled agents to expand and contract along both the direction of motion and laterally.
This allowed simulation of behaviors such as lateral swaying and torso rotation,  which are essential for navigating tight spaces and capturing more nuanced body dynamics in crowd movement.

Additional alternatives to elliptical representations also exist. 
Some researchers have experimented with a ``three-circle body'' approach \cite{Thompson1995a}, which approximates the torso and limbs using overlapping circles.
Others, such as Alonso{-}Marroqu\'{\i}n et al. \cite{Alonso-Marroquin2014}, have employed spheropolygons, geometric shaped with multiple vertices and rounded edges, to more closely mimic the contours of the human body. 

Although point-based or circular approximations can be sufficient at lower densities, the lack of bodily detail becomes problematic as crowd density increases and physical contact intensifies. 
Static 2D shapes, such as circles or ellipses, are inherently limited in their ability to capture complex phenomena like balance recovery and the realistic distribution of forces across the pedestrian's body. 
Although some studies have proposed more anatomically realistic representations, none have fully overcome the challenge of balancing geometric fidelity with computational feasibility in dense-crowd scenarios. This persistent gap highlights the need for a paradigm shift in crowd modeling, one that transcends simplified 2D geometries and embraces the three-dimensional nature of human body, along with the rich physical interactions that emerge at high-density conditions.

\subsection{Microscopic Models}
Microscopic models represent each pedestrian as an individual agent whose behavior emerges from local interactions and decision-making processes.
These models are especially well-suited for capturing fine-grained dynamics, such as collision avoidance, small-group formation, and navigation strategies in buildings.
Within this category, two principal modeling approaches have gained widespread attention: force-based models and velocity-based models.
Each offers distinct assumptions and computational strategies to simulate how pedestrians respond to their environment and to one another.

\subsubsection{Force-based Models}
Force-based models draw inspiration from Newtonian mechanics, representing pedestrians as particles subjected to a superposition of pairwise interaction forces between neighboring individuals.
These interaction mechanisms are derived from diverse anisotropic and nonlinear interaction potentials, which capture both attractive forces - such as those needed to model group behaviors, and repulsive forces, which are essential for collision avoidance and enforcing physical body exclusion.
A pioneering example is the social force model, introduced in the late 1990s by Helbing and Moln\'ar \cite{helbing1995social},  which employs exponential repulsive potentials. This model builds on earlier fundamental work by Hirai and Tarui in the 1970s, who simulated crowd behavior at train stations \cite{Hirai_1975}, as well as the theoretical concepts proposed by Lewin in the 1940s, whose concepts on social fields contributed to understanding group behavior from a psychological perspective \cite{lewin1951field}.

\begin{figure}[!ht]\small
\medskip
\begin{minipage}[c]{.24\textwidth}
\centering Isotropic repulsive interaction\\[0mm]
\includegraphics[width=\textwidth]{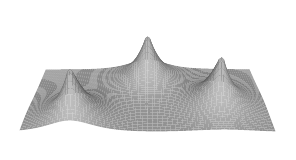}
\end{minipage}
\textcolor{blue}{\large\!\!\!\!$\times$\!\!\!}
\begin{minipage}[c]{.24\textwidth}
\centering Anisotropic\\ effect\\[0mm]
\includegraphics[width=\textwidth]{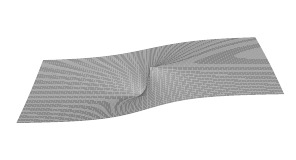}
\end{minipage}
\textcolor{blue}{\large\!$+$\!\!}
\begin{minipage}[c]{.24\textwidth}
\centering Dissipation and\\ input control\\[0mm]
\includegraphics[width=\textwidth]{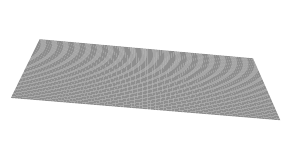}
\end{minipage}
\textcolor{blue}{\large\!\!$=$\!\!}
\begin{minipage}[c]{.24\textwidth}
\centering Force-based\\model\\[0mm]
\includegraphics[width=\textwidth]{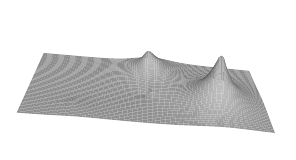}
\end{minipage}\medskip
    \caption{Schematic representation of the force-based models. The repulsion with neighbors is short-range and anisotropic, while the control to the desired direction is a linear relaxation process. Force-based models are the sum of these two components.}
    \label{fig:force-based}
\end{figure}

General force-based models are given by the second-order dynamics \cite{chraibi2011force,Claudia2020}
\begin{equation}
    \ddot{\vec x}_n(t)=\frac1\tau\big[\vec v_0-\dot{\vec x}_n(t)\big]+\sum_{m\ne n}
    \omega\big(\phi_{n,m}(t)\big)\,
    \nabla_{\!\vec x\;}\mathcal U\Big[\Delta \vec x_{n,m}(t),\Delta \dot{\vec x}_{n,m}(t)\Big],
    \label{eq:FB}
\end{equation}
where $\tau>0$ is a reaction (relaxation) time, $\vec v_0\in\mathbb R^2$ is the \emph{desired velocity} coming from a tactical where $\vec x_n(t)$ is the position of the pedestrian $n$ at time $t$, $\tau>0$ is a reaction (relaxation) time, $\vec v_0\in\mathbb R^2$ is the \emph{desired velocity} coming from a tactical modeling level, $\mathcal U\in C^2((\mathbb R^2,\mathbb R^2),\mathbb R^2)$ is an interaction potential that usually depends on the distance to the neighbors - its gradient is the interaction force that can be attractive or repulsive, while $\omega\in C^1(\mathbb R^2,\mathbb R)$ is an anisotropic factor based on the bearing angle $\phi_{n,m}$ that provides more weight to the pedestrians in the direction of motion. 
A schematic representation of the force-based models is given in~\Cref{fig:force-based}. 
For example, in the \emph{Social Force} (SF) model, the interaction force is derived from an exponential repulsive interaction potential based on the distance and the piecewise constant anisotropic factor
\begin{equation}
    \mathcal U_\textsc{sf}(\Delta \vec x)=AB\exp(-|\Delta \vec x|/B),\qquad \omega_\textsc{sf}(\phi)=\left\{\begin{array}{cc}
        1 & \text{if }|\phi|<\kappa, \\
        c & \text{otherwise},
    \end{array}\right.
\end{equation}
where $A,B>0$ are the repulsion range and the characteristic repulsion distance, respectively, while $0<c \ll 1$.
Due to its exponential potential, the interaction in the social force (SF) model is short-ranged. 
In contrast, both the \emph{Centrifugal Force} (CF) model \cite{yu2005centrifugal} and the \emph{Generalized Centrifugal Force} (GCF) model \cite{chraibi2010generalized} feature interactions with an algebraically decaying range. 
Furthermore, the GCF model incorporates a pedestrian shape dependent on speed, modeled as ellipses \cite{chraibi2010generalized}. Unlike the SF model, both the CF and GCF models also consider the relative velocity between pedestrians. 
Today, many extensions of the SF model integrate relative velocity terms and other velocity-dependent mechanisms; see the review in \cite{chen2018social} for further details and references.

Despite their success in reproducing collective phenomena such as lane formation in bidirectional flows, force-based models often rely on simplified two-dimensional geometries (points, circles, ellipses).
This simplification becomes a critical limitation in dense crowd scenarios, where more realistic representations of body shape and physical contact are essential to accurately capture the effects of pushing, falling and balance recovery.
Another challenge lies in models' inherent sensitivity to parameter tuning: Small variations in interaction coefficients can lead to significantly different outcomes.
As a result, although force-based models perform well under moderate-density conditions, they tend to lose accuracy when applied to high-density environments, where a detailed account of physical forces and biomechanical responses becomes indisponsable.

\subsubsection{Velocity-based Models}

Velocity-based models offer an alternative to force-based approaches by focusing on the direct computation of an agent’s desired velocity, rather than simulating forces. These models typically define a set of velocity constraints to ensure collision-free movement, allowing each pedestrian to adjust their velocity based on the anticipated movements of others in the environment. Unlike force-based models, which rely on continuous interaction potentials, velocity-based models are often formulated as optimization or geometric constraint problems, making them particularly well-suited for real-time applications and multiagent systems.

Models such as Reciprocal Velocity Obstacle (RVO) and Optimal Reciprocal Collision Avoidance (ORCA)~\cite{VanDenBerg2008,Arul2021,fiorini1998motion} exemplify this principle by computing admissible velocities that account for the anticipated positions of all neighboring agents.

A general formulation of velocity-based models can be expressed as an optimization problem under constraints \cite{van2020generalized}:
\begin{equation}
    \dot{\vec x}_n(t)=\text{arg}\hspace{-5mm}\min_{\vec v_n\,\not\in\,\cup_{m\ne n}\mathcal{C}_n^m(t)}C_{\vec x(t)}^{\vec v_0}(\vec v_n).
\end{equation}
Here, $\mathcal{C}_n^m(t)$ denotes the collision cone with the neighboring agent $m$, constructed by linearly interpolating the agents' trajectories -- a concept initially introduced for controlling robots amid moving obstacles \cite{fiorini1998motion}.
The cost function $C_{\vec x}^{\vec v_0}$ depends on the desired velocity as well as the positions and velocities of the neighbors.
In its simplest form,  the cost is assumed to be the quadratic deviation from the desired velocity:
\begin{equation}
    \dot{\vec x}_n(t)=\text{arg}\hspace{-5mm}\min_{\vec v_n\,\not\in\,\cup_{m\ne n}\mathcal{C}_n^m(t)}\|\vec v_0-\vec v_n\|^2.
    \label{eq:VO}
\end{equation}
More advanced formulations expand this cost function to account for additional behavioral and biomechanical constraints. For instance, the collision-free model developed in \cite{paris2007pedestrian} incorporates terms associated with speed variations, deviations from preferred speed, and changes in orientation.
Similarly, the RVO model \cite{VanDenBerg2008} extends the quadratic cost with a repulsive potential based on time to collision. 
The ORCA model \cite{van2011reciprocal} retains this quadratic structure, while its extension PORCA  includes an additional velocity-dependent term that prevents the system from freezing \cite{luo2018porca}.

\begin{figure}[!ht]
    \centering
    \input{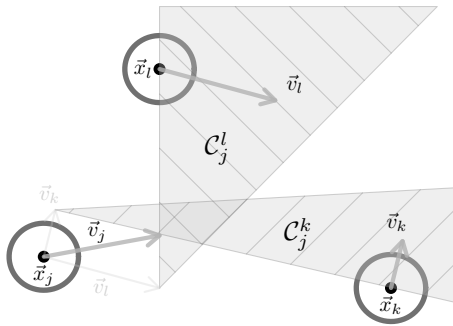}
    \caption{Velocity obstacle cones for an agent as used by RVO-models. No collision occurs when the velocity is set outside the cones.}
    \label{fig:cones}
\end{figure}

By treating pairwise pedestrian interactions as a form of trajectory planning, velocity-based models aim to ensure that agents proactively select paths avoiding collisions while navigating through shared spaces.
For low to moderate densities, these models effectively reproduce collective behaviors -- largely due to social proxemics (the tendency to maintain preferred distances) and anticipatory strategies for collision avoidance (how agents modify their paths to prevent collisions). 
However, as densities increase and more complex physical interactions, such as body contact, force propagation and falling, become dominant, the assumptions of velocity-based models begin to break down.
In high-density scenarios, where physical contact is unavoidable, velocity-based models often struggle to maintain realism or capture critical behaviors like pushing and falling.

Thus, while both force-based and velocity-based models offer valuable insights into pedestrian movement at moderate densities, neither is sufficient for modeling dense crowds, where contact-driven interactions and three-dimensional body dynamics play a pivotal role. Bridging this gap requires novel modeling approaches that incorporate detailed physical processes while remaining flexible across varying density regimes.

\subsection{Macroscopic Models}

Macroscopic models describe crowd dynamics by aggregating individual behaviors into continuous fields such as density, velocity and flow, offering insights at a larger scale compared to microscopic models. These models are particularly useful for simulating large crowds over extended areas, where fine-grained individual interactions are less critical than global movement patterns.
They can be broadly classified into network models, hybrid models, and continuum models \cite{hoogendoorn2015continuum}. 
Network models, which are based on graph theory and heuristic rules, are primarily used to capture pedestrian movement at strategic and tactical levels, such as route choice or facility planning.
In contrast, continuum models treat the crowd as a continuous medium, often analogous to a fluid, and use partial differential equations (PDEs) to describe pedestrian behavior at the operational level, such as local velocity adaptation and density-driven flow.

\subsubsection{Macroscopic Model Classification}

In contrast to microscopic models, which focus on individual pedestrian positions $\vec x_n(t)\in\mathbb R^2$ and velocities $\vec v_n(t)\in\mathbb R^2$ in Lagrangian coordinates $(n,t)$ ($n\in\mathbb N$ is the pedestrian index and $t>0$ the time),
macroscopic models use aggregated variables such as density $\rho(\vec x,t)\in\mathbb R$, flow $\vec q(\vec x,t)\in\mathbb R^2$, and mean velocity $\vec v(\vec x,t)\in\mathbb R^2$ in Eulerian coordinates $(\vec x,t)$. 
Here, $\vec x\in\mathbb R^2$ is the space 
and $t\in[0,\infty)$ is the time. 
Note that density is a scalar field, whereas flow and velocity are vector fields.  
Macroscopic models are often classified by order of their dynamics \cite{cristiani2014multiscale}. 
First-order models rely on the mass conservation equation
\begin{equation}
    \frac{\partial \rho}{\partial t} + \nabla\cdot \vec q(\rho) =0,
    \label{eq:Macro1stOrder}
\end{equation}
where the flux $\vec q(\rho)$ is typically defined as $\vec q(\rho)=:\rho \vec v(\rho)$. 
In practice, the velocity function $v(\rho)=v(\rho)\vec\mu(\rho)$ captures the dependence of pedestrian speed on density, with  $\mu(\rho)$ indicating the desired direction (normalized vector).
The phenomenological scalar relationships between density, flow and the mean speed are called the \emph{fundamental diagram} in traffic engineering. 
There are different forms for the fundamental diagram in the literature (see, e.g., the reviews \cite{seyfried2005fundamental,vanumu2017fundamental}). 
Second-order models add momentum conservation through an additional PED \cite{cristiani2014multiscale}:
\begin{align}
    \left\{~\begin{aligned}
        \frac{\partial \rho}{\partial t} + \nabla\cdot (\rho \vec v) &= 0,\\[1mm]
        \frac{\partial \vec v}{\partial t} + (\vec v\cdot\nabla)\vec v &= \vec a(\rho,\vec v),
    \end{aligned}\right.
    \label{eq:Macro2ndOrder}
\end{align}
The first equation accounts for the conservation of mass (i.e., the pedestrians), while the second equation models the conservation of the momentum. 
The function $\vec a$ in this last equation represents the aggregated \emph{acceleration} of pedestrians. 
It has to depend on the two unknowns $(\rho,\vec v)$ to close the system. 
These models are complemented by the initial conditions $\rho(0,x)=\rho_0(x)$ and $\vec v(0,x)=\vec v_0(x)$ and by the desired direction field $\vec\mu(\rho(x,t))$.

\subsubsection{Main Macroscopic Models}

In the famous first-order model of Hughes \cite{hughes2003flow}, the desired direction field results from an Eikonal equation that adjusts the direction with the density using the gradient of an interaction potential based on the fundamental diagram
\begin{align}
    \label{eq:Eikonal}|\nabla \phi|&=\frac1{v(\rho)},\\
    \vec v(\rho)&=v(\rho)\frac{\nabla \phi}{|\nabla \phi|},\\
     \frac{\partial \rho}{\partial t} + \nabla\cdot \bigl(\rho\vec v(\rho)\bigr) &= 0.
\end{align}
Adding an additional Laplace operator to the squared Eikonal equation \eqref{eq:Eikonal} for a sufficiently small $\varepsilon > 0$ leads to the approximation $\phi_\varepsilon$ of the original solution $\phi$, which is called the viscosity solution \cite{axthelm2016finite} and the nonlinear second-order partial differential equation:
\begin{equation}
    \varepsilon^2\Delta \phi_\varepsilon+\bigl(\nabla \phi\bigr)^2=\frac1{v^2(\rho)}\cdot
\end{equation}
Introducing the Laplacian term as a viscosity solution further improves numerical stability \cite{axthelm2016finite}. 
In the second-order realm, the extension of the one-dimensional Payne-Whitham vehicular traffic model to two-dimensional pedestrian motion \cite{twarogowska2014macroscopic} uses an acceleration function:
\begin{equation}
    a(\rho,\vec v)=\frac1\tau\bigl(\rho v(\rho)\vec\mu - \rho \vec v\bigr)-P(\rho).
    \label{eq:PWped}
\end{equation}
where $P(\rho)= p_0\rho^\gamma$, with $p_0 > 0$, $\gamma > 1$, 
models a repulsive force due to volume filling effect in \cite{twarogowska2014macroscopic}. 
Alternatively, the term $P(\rho)=c^2(\rho)\frac{\nabla\rho}{\rho}$ is used to capture anticipatory effects in \cite{jiang2010higher}.
Some models, like \cite{golas2014continuum}, further include \emph{deviatoric stress} represent to frictional forces.
See \cite{bellomo2011modeling} for an overview of the different model forms.
In all these approaches, the desired direction is often determined by solving the density-based Eikonal equation, sometimes with additional modifications such as the discomfort fields in \cite{jiang2010higher}.

\subsubsection{Mean Field Games}

Some recent macroscopic models combine the tactical and operational levels in a coupled partial differential equation system. 
They are derived from \emph{mean field games} (MFG), introduced by Lasry and Lions \cite{lasry2007mean} and by Huang et al. \cite{huang2006large} in 2006 to describe Nash equilibria in differential games with infinitely many players. 
The first applications to pedestrian dynamics date back to the early 2010s \cite{dogbe2010modeling,lachapelle2011mean,Martin2014}
Pedestrians in a MFG adopt strategies that aim, for instance, to individually minimize their walking time \cite{mazanti2019minimal}. 
In contrast to classical macroscopic models, where the interactions are mainly reactive, the pedestrians are able to anticipate over an arbitrary horizon time. 
In addition, the dynamics are assumed to be stochastic and include a white noise.
The pedestrians are indistinguishable and move according to the same individual optimization criteria. 
The dynamics are considered at the mean field limit over a large number of pedestrians in equilibrium. 
Therefore, pedestrians individually have a minor (infinitesimal) influence on others, and their motions result from the mean-field interaction.
MFG pedestrian models are much less sophisticated than microscopic N-player games.

MFG consists of a set of coupled differential equations. 
The first is a Hamilton-Jacobi-Bellman (HJB) equation, which describes how each pedestrian optimizes its trajectory over time as a feedback (backward) equation. 
The second is a Fokker-Planck (FP) equation that describes how the distribution of pedestrians (i.e., the density) evolves over time under the influence of collective behavior (forward equation). 
A general formulation is given by \cite{cristiani2014multiscale}
\begin{equation}
    \left\{~\begin{aligned}
&-\frac{\partial u}{\partial t} + H(x,t,\nabla u,\rho)=\frac{\sigma^2}2\Delta u,\\[0mm]
&\frac{\partial \rho}{\partial t} -\nabla\cdot\biggl(\rho\frac{\partial H}{\partial p}(x,t,\nabla u,\rho)\biggr) = \frac{\sigma^2}2\Delta \rho,
\end{aligned}\right.
\end{equation}
where $u$ is the value function representing the minimum cost and $H$ is the Hamiltonian representing an optimal trade-off between control effort and system dynamics via an interaction potential. 
In the HJB equation the value function is derived from the Hamiltonian, while the FP equation describes the evolution of the density according to a continuity equation where the velocity is given by the optimal drift $-\partial H/\partial p$, with $p=\nabla u$, corresponding to a local fundamental diagram. 
In both equations, the noise plays the role of a second-order diffusion.

The MFG is derived in the mean-field limit from the general microscopic dynamics given by the stochastic differential equation:
\begin{equation}
\mathrm dx(t)=b(\alpha(x,t),\rho(x,t))dt+\sigma \mathrm d W(t).
\end{equation}
where $b(t,x,\alpha,m)$ is the drift term derived from the interaction potential, $\alpha$ is the optimal velocity control, while $W_n$ is a standard Brownian motion with $\sigma$ as diffusion coefficient (noise amplitude). 
Quadratic mean-field games are specific tractable cases for which the interaction potential is quadratic  \cite{ullmo2019quadratic,butano2024mean}, i.e. the drifts in the microscopic dynamics are linear. 
Quadratic mean-field games are given by
\begin{equation}
    \left\{~\begin{aligned}
&-\frac{\partial u}{\partial t} + \frac12|\nabla u|^2 - U(\rho(x)) =\frac{\sigma^2}2\Delta u,\\[0mm]
&\frac{\partial \rho}{\partial t} -\nabla\cdot(\rho\nabla u) = \frac{\sigma^2}2\Delta \rho.
\end{aligned}\right.
\end{equation}
The term $-\nabla\cdot(\rho\nabla u)$ represents the movement of the crowd following optimal paths, while $U$ is an increasing function that penalizes high pedestrian densities. 
Quadratic mean-field games provide analytical tractability and explicit solutions \cite{toumi2020tractable}. Additionally, the HJB-FP system that governs pedestrian dynamics can be efficiently solved using fast finite-difference methods, further enhancing scalability. The presence of quadratic terms also allows for the application of Riccati equations, streamlining optimization and making the approach well suited to large-scale simulations with low or medium density levels.

\subsection{Limitation of Actual Models for Dense Crowds}

In summary, classical microscopic and macroscopic pedestrian models perform adequately for low and intermediate density levels, where interactions are primarily governed by proxemics and collision-avoidance behavior and where the concept of fundamental diagram applies.
However, these models are limited when applied to dense situations where the interactions are dominated by physical pushing forces, body compression, and wedging dynamics. 
While some models can partially reproduce certain collective behaviors of dense crowds by incorporating additional parameters and mechanisms, there is no universal model or consensus.
In extreme crowd conditions, individuals can become so tightly packed that their bodies effectively lock together, preventing movement, a phenomenon we refer to as wedging. This process occurs under circumstances where collision avoidance becomes impossible, and the normal regulation or optimization of velocity based on neighbor distance and displacement no longer applies.

The current modeling paradigms, be they microscopic models based on pairwise interactions or macroscopic models grounded in continuum theory, struggle to capture these complex physical interactions. Concepts such as fundamental diagram, collision avoidance, and trajectory optimization lose their applicability in such scenarios, as the primary concern shifts to managing the accumulation of compression forces and ensuring balance recovery. In addition, existing models do not adequately account for the conditions under which wedging occurs, nor do they offer insights into strategies that might prevent these dangerous configurations.

While some ad-hoc macroscopic models with additional parameters can reproduce some collective behavior observed in dense crowds, they lack the fine-scale modeling of pedestrian bodies in 3D and understanding of the microscopic and sub-microscopic mechanisms that take place at high densities, such as balance recovery. 
The coupling of such individual-related features with consistent macroscopic derivations capable of describing the observed collective phenomena that arise in dense crowds remains to be developed.
To overcome these limitations, new models must be developed that integrate detailed three-dimensional representations of pedestrian bodies in a consistent multiscale framework. 

\section{Biomechanics and Physical Interactions in Dense Crowds}
\begin{figure}[!ht]
    \centering
    \includegraphics[width=1\linewidth]{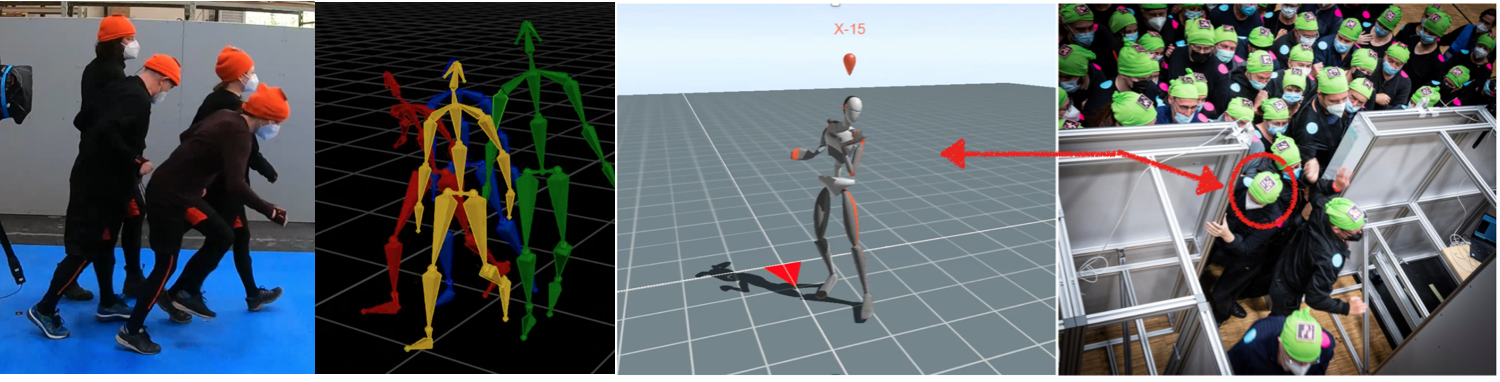}
    \caption{Experimental evacuation from narrow bottlenecks. Pedestrians use different parts of their body to
squeeze through and evacuate. The inability of state-of-the-art models to simulate this competitive evacuation
process stems from their reliance on underlying 2D shapes and from lack of empirical understanding of these mechanisms.}
    \label{fig:3D-scenarios}
\end{figure}

Building on our previous discussion of crowd modeling, recent research has increasingly focused on biomechanics and physical interactions that occur
in dense formations. 
This emerging line of research is at the crossroads of established research on standing balance and the study of collective dynamics. 
On one hand, extensive work has been conducted within research communities dedicated to standing balance and balance recovery in individuals under various conditions and across different populations. 
On the other hand, researchers in collective dynamics are now trying to understand how standing balance is affected in crowded environments, where unique physiological factors and complex motion constraints come into play.
\Cref{fig:3D-scenarios} shows in experimental setups how pedestrians try to maintain their balance in challenging and crowded scenarios.

To fully grasp how dangerous dense crowds can be, it is fundamental to understand how the human body responds to physical interactions in such contexts. 
At the human body level, two main mechanisms emerge in response to these physical interactions in dense crowds. The first is the use of whole-body movements to prevent falls. 
These balance-recovery movements, which encompass coordinated actions that involve the feet, legs, upper body, arms and hands \cite{Feldmann_2024_results2,Chatagnon_2024}, are essential to maintain stability.
The second mechanism is related to the mechanical properties of body tissues; in densely packed crowds, the upper body experiences significant compression, leading to discomfort, injury, or even death \cite{Sieben_2023}.
In light of these observations, it becomes important to understand 
how individuals leverage balance recovery movements to avoid falls.

In this section, we first review the methods and experimental paradigms that have been developed to understand standing balance and balance recovery following external perturbations on single individuals. 
We then present a review of the existing literature on the experimental study and modeling of physical interactions and balance recovery in dense crowds.
This dual perspective not only aims at enlarging our understanding of individual biomechanics but also informs the development of more realistic crowd models aimed at enhancing safety in high-density situations.

\subsection{Standing Balance and Recovery Strategies of Single Individuals}

In 2019, falls were estimated to be the second leading cause of unintentional injury or death worldwide, with more than 684,000 people involved in fatal fall incidents \cite{WHO_2021}. Falls are a major public health problem, particularly in the elderly. Consequently, research on standing balance has become extensive, particularly in relation to responses to various types of internal and external perturbations \cite{Vinik_2017,Winser_2019,Mikos_2021,Johnson_2023}.

We focus here on studies regarding reactions to external perturbations, as these are inherently more relevant to the physical interactions encountered in dense crowds.
Perturbations in standing balance can arise in a wide variety of forms, such as mechanical pushes, surface instability, or unexpected shifts in posture, and are often investigated through parameter studies. 
This approach allows researchers to isolate and assess the influence of specific factors, such as age, health conditions, or specific type of terrain, on the balance recovery and stability responses \cite{Mangold_1996,Schulz_2006,Watier_2023}. 

\subsubsection{Methods and Metrics}

To understand how individuals fall in crowds, we need to define the \emph{Limit of Standing Balance} (LoSB) or limit of stability \cite{Horak_2006}. This is the limit beyond which individuals must initiate recovery strategies involving \emph{Change-in-support} to regain balance following perturbations. This \emph{Change-in-support} corresponds to a modification of the \emph{Base of Support} (BoS), which is defined as the area that includes all points of contact between individuals and their support surfaces. Modifications of the base of support may include step initiations and physical interactions with the surrounding environment, including other individuals \cite{Maki_1997}.

One of the most important variables used to study standing balance, and human motion in general, is the position of the Center of Mass (CoM). This can be defined as the average position of the body mass distribution in space.
In early work, the moment when the projection of the CoM on the ground reaches the boundary of the Base of Support (BoS) was suggested as the functional LoSB \cite{Shumway-Cook_1995,Winter_1995}. However, this proposition is only limited to static situations, and individuals tend to initiate recovery strategies before reaching this limit in dynamic scenarios \cite{Hof_2005}. To create a more general model about the LoSB new metrics have been created such as \emph{Zero Moment Point}, originally proposed for robotic applications. Other simpler metrics may also be used to study the LoSB. In \cite{Hof_2005}, Hof et al proposed the use of an inverted pendulum model to represent the standing individual and they derived the concept of \emph{Extrapolated Center of Mass,} (XCoM) which proved to be an efficient tool to access standing balance. Another approach consists in using the \emph{Time to Boundary,} that corresponds to the time required by the CoM to exit the BoS \cite{Schulz_2006}. Using this simple yet effective method, the initiation of steps following perturbations in the sagittal plane could be archived with an accuracy reaching $80\%$. 
Later, this accuracy was improved using trained neural network models based on kinematic features \cite{Emmens_2020}.

In addition to the previous quantities, the angular momentum and its time derivative can also be studied. These variables have  been studied mainly in walking subjects and have been shown to increase during unstable gait phases \cite{Herr_2007}. Angular momentum can also be used to identify the contribution of body parts to balance control or to study the first steps after  standing upright \cite{Begue_2021,Watier_2023}. 

\subsubsection{Experimental Paradigms}

In terms of experimental paradigms, balance recovery has been studied for a variety of situations \cite{Tokur_2020}. In a crowd, perturbations can occur while other voluntary movements are also ongoing, e.g., walking towards a target. However, we are focusing here on dense crowds, in which only a limited range of motion is possible  \cite{Sieben_2023}. Therefore, we restrict our attention here to balance recovery of static standing postures following external perturbations.
 We can define, by analogy to physics, the hypothesis of a quasi-static dense crowd. With this hypothesis, we consider that body motion is still possible (e.g., for balance recovery) but no target-oriented motion takes place. 
 One may also note that balance recovery could be studied for very slow locomotion as well \cite{Mierlo_2023}, as this may occur in dense crowds \cite{Mierlo_2023}.
Studies relative to standing balance recovery following external perturbations can be grouped relative to the two main families of perturbations applied in experimental protocols. 

The first type of perturbations are noncompliant perturbations. For this type of perturbations, an unbalanced posture is imposed on the participants’ body. Non-compliant perturbations can be obtained using a \emph{tether-release} method, i.e. releasing a cable which maintains an unbalanced forward leaning angle position \cite{Thelen_1997,Carty_2011}. Another way of applying non-compliant perturbations is to directly impose body displacement and velocity using a velocity-controlled mechanism \cite{Mille_2003}.

Compliant perturbations can also be investigated. For this kind of perturbation, the response of the subjects can modify the body displacement induced by the perturbation. These external perturbations can be obtained by generating perturbations through different techniques such as; moving ground platforms \cite{Inkol_2018,Batcir_2022} or force-controlled perturbations \cite{Schulz_2006,Robert_2018,Emmens_2020,Zelei_2021}. Regarding the directions of the perturbation, most of the above-mentioned studies focuse only on the perturbation in the anteroposterior and mediolateral directions. Only a few studies have investigated intermediate perturbation directions using the moving ground paradigm \cite{Moore_1988,Maki_1997} or force-controlled perturbations \cite{Chatagnon_2023}. One must keep in mind that across all studies only a limited number of perturbations can be investigated, while perturbations in a real crowd context can arise from all possible directions, leading to different recovery strategies \cite{Chatagnon_these}.

\subsection{Collective Dynamics with Physical Interactions}

The current section is dedicated to research that specifically addresses collective motion induced by physical interactions. We attempt to provide an overview of recent studies on physical interactions and balance recovery in human crowds. In particular, we review the different experimental setups and associated results on physical interaction and standing balance in dense crowds. We then focus on the models and numerical paradigms that have been proposed to represent an analysis of these specific crowd configurations.

\subsubsection{Field Studies}

Many studies have documented accidents occurring in dense crowds (see, e.g., \cite{Zhou_2017,Sieben_2023,Son2025} or \cite{feliciani2023trends} and references therein). 
However, only a few studies include quantitative analysis of the crowd and its dynamics. 
A pioneering analysis of \emph{Jamaraat Bridge} during the Hajj pilgrimage to Mecca in Saudi Arabia has shown that dense crowds can describe large stop-and-go waves and show multidirectional motions \cite{helbing2007dynamics,Helbing2009}. 
A study of the Love Parade in Duisburg, Germany (2010), has highlighted the phenomenon of crowd quakes \cite{ma2013new}. 
Specific observations at crowded concerts and festivals also show coordinated dynamics such as wave phenomena, collective oscillations and vortices \cite{silverberg2013collective,bottinelli2016emergent,Bottinelli_2018}. 

\begin{figure}[!ht]
    \centering
    \includegraphics[width=0.8\linewidth]{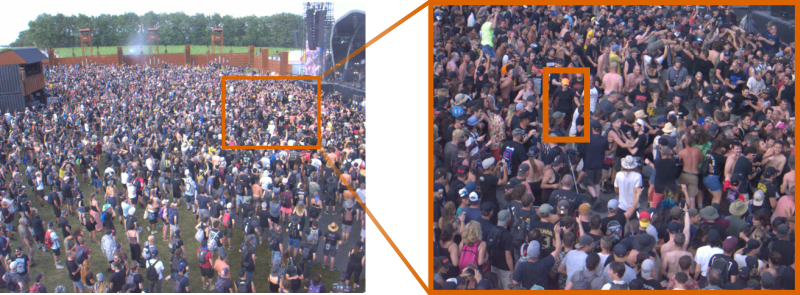}
    \caption{ Bird's eye view image highlighting the position of a participant wearing a motion capture suit while immersed in the crowd at a punk rock concert. Figure from \cite{Chatagnon_these}.}
    \label{fig:hellfest_exp}
\end{figure}

Recently, Gu and co-authors have used advanced computer vision techniques to collect large datasets of massive crowds that repeat every year during the San Firm\`in festival in Pamplona, Spain \cite{gu2025emergence}.  
Data from \emph{Plaza Consistorial}, which is approximately 50 m long and 20 m wide, and where up to 6 pedestrians per square meter (locally up to 9) are regrouped in a confined space before the start of the festival.
The data collection includes a fine-grained representation of the density and velocity field sequences, allowing a precise analysis of the dense crowd dynamics.
The results reveal the organization of the crowd into large chiral oscillations when the density exceeds a critical threshold (about 4 pedestrians per square meter), coordinating the orbital motion of hundreds of individuals without external guidance. See~\Cref{fig:hellfest_exp}.

Recent efforts also include the MADRAS project, which provides one of the first large-scale field datasets capturing dense pedestrian dynamics in real-world conditions, covering densities up to 4 pedestrians per square meter. 
Collected during the 2022 Festival of Lights in Lyon, France, the data include macroscopic crowd flows, GPS traces, contact statistics, and nearly 7000 microscopic trajectories, offering a unique opportunity to benchmark and calibrate crowd models under realistic conditions \cite{Dufour2025}.

One of the reasons field studies are rare is the difficulty in collecting data from dense crowds.  
However, the advencement of modern computer vision techniques makes it possible to accurately measure density and velocity fields, or even to track individuals.  
Nevertheless, the collection of submicroscopic pedestrian characteristics such as pedestrian positions, is not possible in field studies and must be investigated using different methods.


\subsubsection{Laboratory Experiments}
In order to deal with the limitation of in-situ observation, recent experimental paradigms have been created allowing to study collective dynamic due to physical interaction in the laboratory. These experiments only consider a limited number of participants, thus do not allow replicating large-scale motion observed during a real life scenario. 
However, a greater variety of measurement techniques can be used in such controlled experimental setups. 
This provides access to higher quality measurements of head trajectories as well as lower scale measurements such as whole body motion recording and contact force measurements.

\begin{figure}[!ht]
    \centering
    \includegraphics[width=0.8\linewidth]{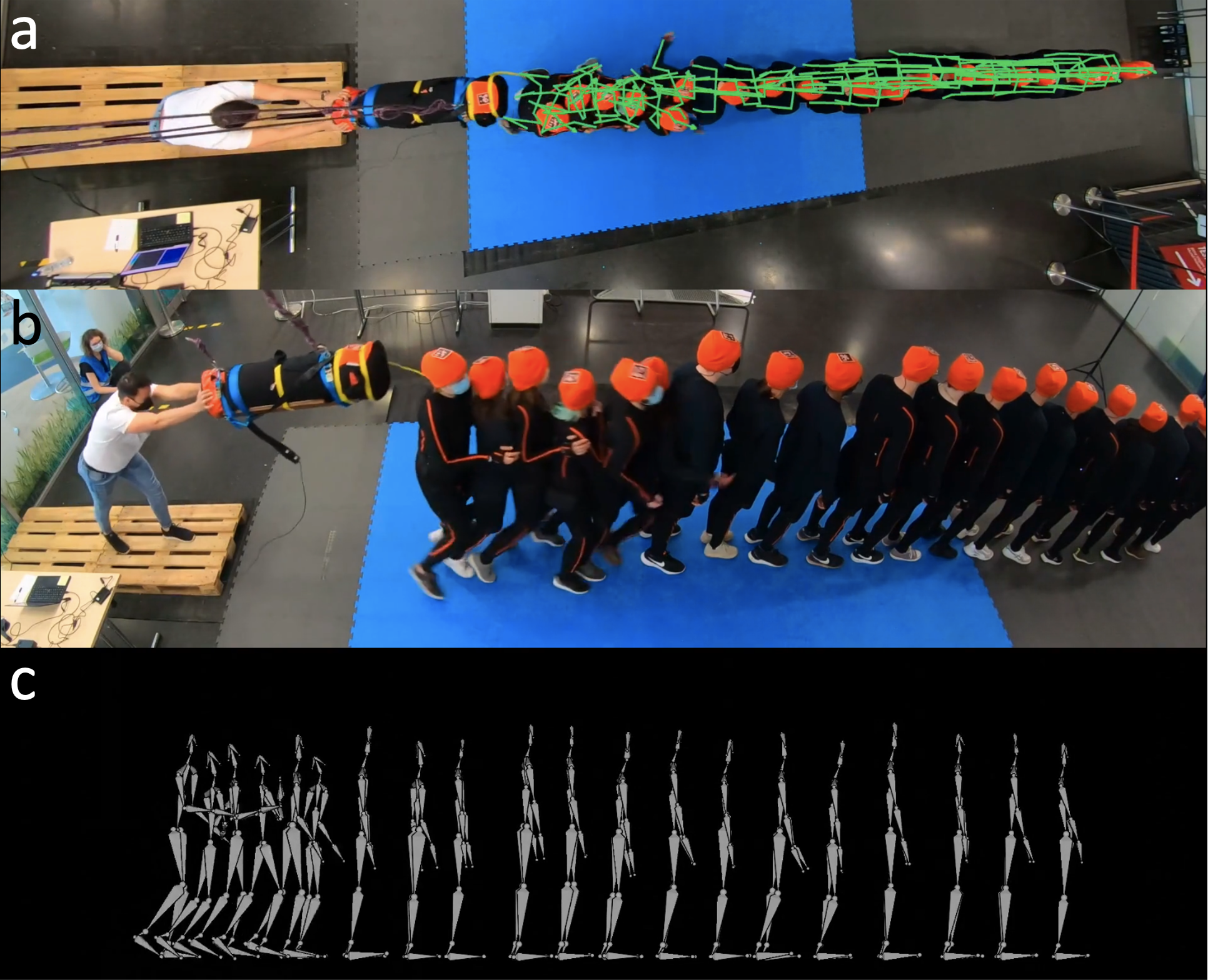}
    \caption{Impulse propagation through a queue of 20 participants. (a) Top view of the experiments, allowing head tracking of the participants. (b) side view of the experimental set-up. (c) reconstructed motion-captured data with accurate relative placement of participants based on head tracking trajectory data \cite{Feldmann_2024_data}.}
    \label{fig:balance_recovery}
\end{figure}

One of the first experimental studies to propose such an approach is the work of \cite{Wang_2018} to study propagation of external perturbation along a single queue of participants. This experimental setup was also investigated by \cite{Feldmann_2023} to investigate the propagation of force-controlled perturbation in a row of participants. Later, this experimental paradigm was broadened using a larger number of participants as well as different dense group formations and perturbation methods. \Cref{fig:balance_recovery} illustrate the experimental setup as well as the reconstructed motion-captured data recorded during the experiments.

The dataset resulting from this experiment is, to our knowledge, the largest dataset on balance recovery following external perturbation in dense crowd formations with full-body motion recording of participants \cite{dataset_ped_2022_6}. These experiments led to several findings especially regarding the propagation of the perturbation in such crowd formation \cite{Feldmann_2024_results1}, the different reaction stages of the participants after perturbation in such a setup \cite{Feldmann_2024_results2} and the modification of individual recovery strategies observed to maintain balance in dense crowd environments \cite{Chatagnon_2024}. 

Regarding stepping strategies following external perturbation, one could also refer to the work of \cite{Li_2023}
who investigated step characteristics in a two-person queue setup. However, this study is very limited as only three participants were involved.

More and more studies also begun to investigate the force constraints undergone by the individual in either dense static \cite{Wang_2020} or dynamic \cite{Li_2020} group situations. Such measures are usually obtained using  pressure measurement pads. 
One of the latest studies by Shen et al. \cite{Shen_2024} investigated the impulse propagation from a participant to a static obstacle following external perturbations. In this study, the authors could draw a link between impulse propagation and the recovery step taken by the participants. 

Finally, a new experimental paradigm also emerges to study physical interaction in crowded environments. As illustrated in, Chatagnon et al. proposed for example the use of IMU-based motion capture to record balance recovery of participants following physical interaction in crowds of punk rock concerts \cite{Chatagnon_these}. 
This kind of experimental paradigm is at the crossroads between field observation and controlled laboratory experiments. This provides a better ecological validity while still allowing some control over the experimental setup. 
However, these types of experiments are still extremely difficult to create on larger scales and require strong collaboration between event organizers and researchers.

\subsubsection{Models and Simulations}

A number of models have been created following observation of large-scale field studies \cite{Gu2025} and smaller scale laboratory experiments, such as the ``human domino process'' to represent the propagation of physical interaction in a queue of individuals \cite{Wang_2019}. However, these models are often limited to only representing crowd motion in specific contexts, similar to the context in which the experimental observations occurred. In addition, high-level features of dense crowds can generally be replicated in simulation using both macroscopic \cite{golas2014continuum} and microscopic \cite{VanToll_2021} crowd representation paradigms. 
Several second-order macroscopic models with additional parameters and mechanisms operating only at high densities exist in the literature \cite{yu2007modeling,silverberg2013collective,golas2014continuum}.
Microscopic models can also incorporate additional parameters in case of contact \cite{helbing2000simulating}. 
However, sub-microscopic dynamics (i.e., at an individual's limb level) observed during laboratory experiments requires a much finer representation of the body shape.

The importance of human body shape for crowd simulation was suggested in the early works of Thompson and Marchant in \cite{Thompson_1995}. 
However, the representation of disc-like agents remained dominant up to now because of its attractive simplicity. 
This representation paradigm is nevertheless very limited to represent the complexity of physical interactions and standing balance \cite{Kim2015}. 
To address these limitations, a new representation paradigm for human body has been developed featuring simple limb representation \cite{Shang_2024}. 
This representation paradigm could then be used in physics-based simulation to evaluate the recovery ability of standing agents \cite{Jensen_2023}. 
However, multiple challenges are yet to be tackled in order to integrate accurate representation of physical interactions in dense crowds, such as limbs contact detection \cite{Gomez_2024}. 
Once may also keep in mind that this more complex representation of human body shapes also come with significant computational cost. 
Solutions must be found to enable these methods to be used to simulate large-scale events, for which risk assessment of dense crowds is the most critical.

\subsection{Discussions}

This section has outlined the ongoing advancement of knowledge aimed at understanding the physical interactions that govern dynamics in dense crowds.
On the one hand, field observations have informed the development of models capable of describing large-scale dynamics of dense crowds. 
On the other hand, controlled laboratory experiments have enabled for a more detailed analysis of
sub-microscopic dynamics, such as contact forces and limb motion, associated with physical interactions.
Recent experimental findings are increasingly being integrated into the development of accurate simulations of dense crowds. However, further efforts are needed to advance this line of research by creating new experimental and simulation paradigms that better capture the complexity of such environments.  
The study of dense crowds and the physical interactions inherent in these contexts must continue to gain momentum to enable concrete applications in accident prediction and risk assessment for major events. 

To conclude this section, we would like to highlight some thoughts of research directions that we believe should be explored in order to draw a complete picture of dense human crowds.

First, we observed that physical interactions between individuals in crowds can be conceptually divided into two main aspects, the \emph{actions} that generate external perturbations and the \emph{reactions} of one or a set of individuals. 
Responses, also known as balance recovery motions, may include step recovery strategies, momentum damping mechanisms, and even follow-up actions on other individuals. This last type of recovery can cascade into dangerous collective movements, also known as crowd collapses \cite{Zhou_2017}.
To the best of our knowledge, physical interaction has so far been studied mainly from the point of view of \emph{reactions}, by examining how individuals react to a given set of external perturbations, taking into account a given set of influencing factors (e.g. age, terrain, sensory conditions). 
We are now missing fundamental information regarding the \emph{actions} from which these external perturbations originate. 
To fill this gap, new measurement techniques and novel experimental paradigms must be developed to quantitatively assess the strength, location and temporal evolution of physical contacts within dense crowds.

In addition, while recovery from external perturbations is crucial to understanding how individuals maintain standing balance, in the most critical scenario, wedging effects can be observed and movement is simply no longer possible\footnote{The authors are not aware of any published work quantifying the movement threshold in wedged conditions and call for further research in this area.}
In these dense crowd scenarios, static constraints can result in severe outcomes such as chest compression and fainting \cite{Sieben_2023}. 
In light of these observations, we argue that understanding the physiological limits of the human body under compression is  essential. While some studies have explored human tolerance to compressive forces \cite{kroll2017acute}, they remain underutilized in pedestrian dynamics research. Future work should integrate findings from biomechanics and medical sciences to better identify critical thresholds and safety margins in high-density environments \cite{Wang_2020}.

Finally, physical interactions are intrinsically related to behavioral and psychological changes that arise in high-density conditions. These changes may themselves be caused by physical discomfort, while also contributing to further crowding and contact. For example, motivation and perceived urgency are known to vary with density, influencing individual movement decisions \cite{Luegering_2023}.

Understanding the interplay between psychological state and physical behavior is crucial for a complete description of dense crowd dynamics. Although a comprehensive review is beyond the scope of this work, recent research in this domain is rapidly evolving \cite{Templeton_2015,Adrian_2020, Uesten_2023,Sieben_2025}, with practical implications in behavioral detection related to physical contact and distress \cite{Alia_2024}.

\section{Conclusion}

In this chapter, we have identified several challenges in modeling dense crowd dynamics. 
First, we began by reviewing a range of existing models for crowd dynamics. From microscopic approaches that detail individual trajectories featuring body representations to macroscopic frameworks that describe aggregated quantities like density and flow.

Although these models have been instrumental in capturing pedestrian behavior at low to moderate densities, they fall short in high-density scenarios, where physical interactions dominate. The limitations of traditional 2D disk-like representations become particularly evident when trying to simulate phenomena such as wedging, balance recovery, and the complex transmission of contact forces that emerge at high crowd densities.

We then explored alternative modeling paradigms,  drawing on insights from disciplines, such as biomechanics. 
In the second section of this chapter, we have reviewed various methodologies and techniques that can be used to study physical interactions and balance recovery mechanisms following external perturbations. These methods, combined with empirical observations from field studies and laboratory experiments, provide a foundation for new investigations and modeling paradigms.

To address the underlying complexities of dense crowds, many studies on human crowds have emerged from collaborations between researchers with multiple backgrounds such as social and computer sciences, physics, mathematics, or biomechanics. 

In addition, recent advances in computer vision through machine learning, as well as improvements in of IMU-based motion capture, offer new possibilities to collect accurate field measurements in dense crowd situations.

A key challenge moving forward is the development of a unified modeling framework that integrates insights from diverse research domains. Such a framework  must combine two essential components: large-scale density dynamics and detailed representations of individual pedestrian body shapes. By achieving this, models can effectively capture emergent phenomena that lead to critical situations while accurately depicting contact forces and stress distributions.

A unified model should allow smooth transitions between low- and high-density regimes, ensuring that the mechanisms governing collision avoidance at low densities seamlessly give way to the detailed physical interactions required at high densities. Achieving this continuity  would not only provide a more comprehensive understanding of pedestrian dynamics across different crowd conditions but also improve predictive capabilities and improve safety management strategies for large-scale events.

There is also considerable promise in the integration of data-driven techniques with traditional physics-based approaches. 
By leveraging large-scale 3D motion databases and state-of-the-art deep learning algorithms, future research can refine model parameters, validate simulations against real-world observations, and ultimately predict critical events with higher accuracy.

In summary, building robust and predictive models of dense crowd dynamics requires bridging the gap between detailed experimental observations and theoretical modeling, with a particular emphasis on  seamless transitions between low-density collision avoidance and high-density physical contact. 
This will involve not only refining the representation of pedestrian bodies and their interactions but also embracing interdisciplinary approaches that draw from biomechanics, data science, and complex systems' theory. 
Such integrative efforts hold the potential to  significantly advance our understanding of crowd behavior and, more importantly, inform practical strategies for crowd management and accident prevention in large-scale events.


\end{document}